\documentclass[aps,prl,groupedaddress,longbibliography,twocolumn,showkeys,showpacs]{revtex4-2}
\usepackage{commath}
\usepackage{mathtools}
\usepackage{amssymb,amsfonts}
\usepackage{amsmath}
\usepackage{hyperref}
\usepackage{graphicx}
\usepackage{natbib}
\usepackage{dutchcal}
\usepackage[usenames,dvipsnames]{xcolor}
\graphicspath{{Figures/}}
\begin{document}
\title{Unruh effect and quantum entanglement for the non-uniform Rindler spacetime}

\author{Manuel de Atocha Rodr\'iguez Fern\'andez}
  \email{mda.rzfz@gmail.com}
\affiliation{Departamento de F{\'\i}sica, CUCEI, Universidad de Guadalajara, Av. Revoluci\'on 1500, Guadalajara, CP 44430, Jalisco, M\'exico}

\author{Alexander I. Nesterov}
  \thanks{Deceased}
\affiliation{Departamento de F{\'\i}sica, CUCEI, Universidad de Guadalajara, Av. Revoluci\'on 1500, Guadalajara, CP 44430, Jalisco, M\'exico}

\author{Gennady P. Berman}
  \thanks{Deceased}
\affiliation{Theoretical Division, T-4, Los Alamos National Laboratory, Los Alamos, NM 87544, USA}

\author{C. Moreno-González}
\affiliation{Departamento de F{\'\i}sica, CUCEI, Universidad de Guadalajara, Av. Revoluci\'on 1500, Guadalajara, CP 44430, Jalisco, M\'exico}

\date{\today}

\begin{abstract}
While the Unruh effect has traditionally been studied under the assumption of uniform acceleration, a simplification motivated by experimental considerations, it is not necessarily true for all non-inertial motions. We propose a novel approach for the indirect detection of the Unruh effect without relying on the former restriction. Previous studies have shown that probing the decoherence of an Unruh–DeWitt detector can significantly reduce the acceleration required for observing the effect by several orders of magnitude compared to earlier proposals. Building on this idea, we develop a theoretical framework describing a non-inertial observer equipped with a detector undergoing non-uniform, time-dependent acceleration. We show that, in a non-uniformly accelerated Rindler spacetime, the particle distribution perceived in the Minkowski vacuum acquires a time-dependent modification of the standard Unruh spectrum. Furthermore, we demonstrate that the inclusion of quantum entanglement leads to a deformation of the Minkowski vacuum into squeezed states. \\
Throughout this work, natural units are employed: $\hbar = c = k_B = 1$.
\end{abstract}

%\pacs{}

\keywords{Unruh effect, non-uniform accelerated frames, quantum decoherence, quantum entanglement.}

\preprint{Draft. Current version: V1}

\maketitle
%%%%%%%%%%%%%%%%%%%%%%%%%%%%%%%%%%%%%%%%%%%%%%%%%%%%%%%%%%%%%%%%%%%%%%%
%%%%%%%%%%%%%%%%%%%%%%%%%%%%%%%%%%%%%%%%%%%%%%%%%%%%%%%%%%%%%%%%%%%%%%%
%%%%%%%%%%%%%%%%%%%%%%%%%%%%%%%%%%%%%%%%%%%%%%%%%%%%%%%%%%%%%%%%%%%%%%%
\section{Introduction}
%%%%%%%%%%%%%%%%%%%%%%%%%%%%%%%%%%%%%%%%%%%%%%%%%%%%%%%%%%%%%%%%%%%%%%%
\subsection{Uniformly accelerated observers}
Consider an observer moving with uniform acceleration $a$ (that is, a Rindler reference frame) in the $z$-direction with respect to an inertial frame in the Minkowski spacetime. The transformations from the coordinates $(t, z)$ in the Minkowski spacetime to the Rindler coordinates $(\tau, \zeta)$ are given by (\cite{Rindler_1966zz}, \cite{Misner1973}, \cite{Moller2011}):
\begin{align}
\label{EQ01}
t = & \frac{1 + a\zeta}{a} \sinh(a\tau), \\
z = & \frac{1 + a\zeta}{a} \cosh(a\tau).
\label{EQ02}
\end{align}
The metric in the Minkowski spacetime can be written as $ds^2 = dt^2 - dz^2$. Utilizing these coordinate transformations, we can be determine that the metric in Rindler spacetime is given by:
\begin{align}
ds^2 = (1 + a\zeta)^2 d\tau^2 - d\zeta^2.
\label{EQ03}
\end{align}
%%%%%%%%%%%%%%%%%%%%%%%%%%%%%%%%%%%%%%%%%%%%%%%%%%%%%%%%%%%%%%%%%%%%%%%
\subsection{Non-uniformly accelerated observers}
\noindent For an observer moving with time dependent acceleration, the coordinate transformation from the rest reference frame (Minkowski spacetime) to the non-uniformly accelerated frame can be written as (\cite{Misner1973}, \cite{Moller2011}, \cite{Kasner_1925}):
\begin{align}
\label{EQ04}
t = & \int d\tau \cosh{\chi(\tau)} + \zeta \sinh{\chi(\tau)}, \\
z = & \int d\tau \sinh{\chi(\tau)} + \zeta \cosh{\chi(\tau)}.
\label{EQ05}
\end{align}
Here, ${\chi}(\tau)$ is the non-uniform speed of the observer. From here it follows that the non-uniform Rindler metric is:
\begin{align}
ds^2 = \big( 1 + \alpha(\tau)\zeta \big)^2 d\tau^2 - d\zeta^2,
\label{EQ06} 
\end{align}
where $\alpha(\tau) = \dot{\chi}(\tau)$ is the non-uniform acceleration of the observer and the dot denotes derivation with respect to $\tau$.

\begin{figure}
\begin{center}
\scalebox{0.55}{\includegraphics{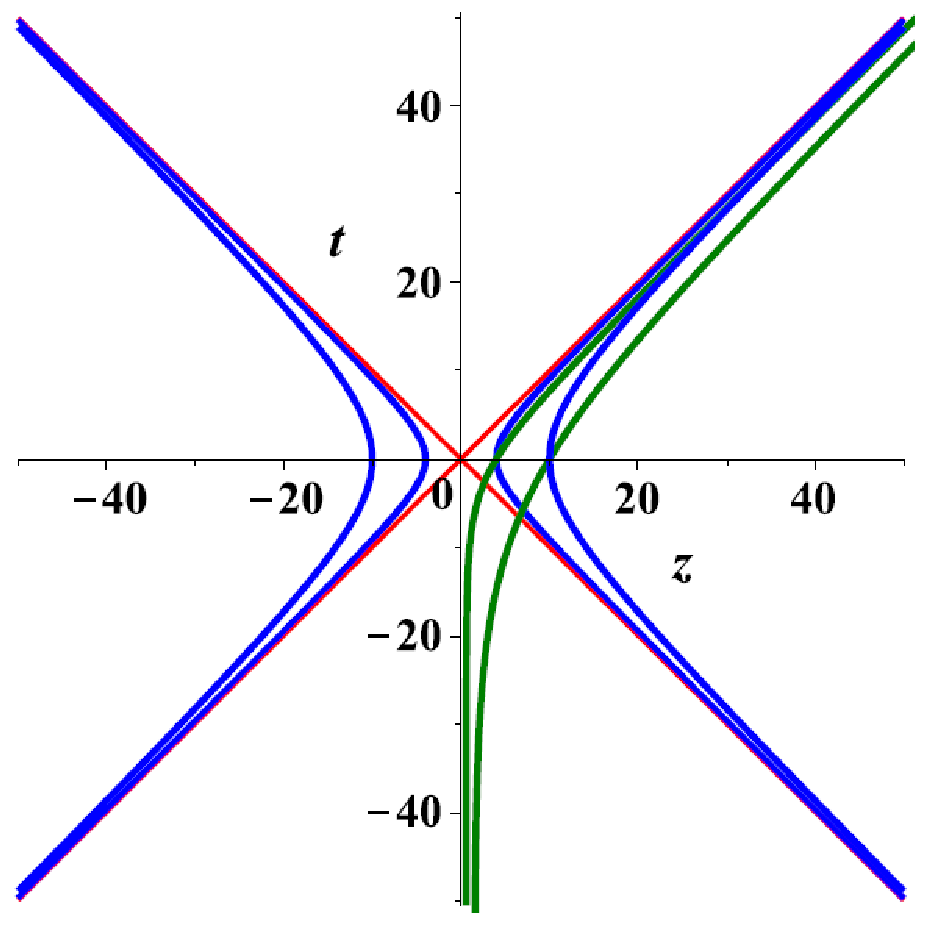}}
\end{center}
\caption{The non-uniform Rindler spacetime $\mathbf{nuR}$ is a non-inertial construction for an observer moving with non-uniform acceleration in the z-direction. In this $\rm{z}$-$\rm{t}$ plane, the red lines are the event horizons; the blue lines are the Rindler trajectories; and the green lines denote the Rindler-generalized trajectories.
\label{Fig1}}
\end{figure}

%%%%%%%%%%%%%%%%%%%%%%%%%%%%%%%%%%%%%%%%%%%%%%%%%%%%%%%%%%%%%%%%%%%%%%%
%%%%%%%%%%%%%%%%%%%%%%%%%%%%%%%%%%%%%%%%%%%%%%%%%%%%%%%%%%%%%%%%%%%%%%%
%%%%%%%%%%%%%%%%%%%%%%%%%%%%%%%%%%%%%%%%%%%%%%%%%%%%%%%%%%%%%%%%%%%%%%%
\section{Quantization of the scalar field}
%%%%%%%%%%%%%%%%%%%%%%%%%%%%%%%%%%%%%%%%%%%%%%%%%%%%%%%%%%%%%%%%%%%%%%%
\subsection{Rindler spacetime: conformally flat metric}
Consider an observer moving in the Rindler spacetime. To perform the quantization of a scalar field, it is convenient to express the lightcone coordinates as follows:
\begin{align}
\tilde{u} = t - z = & -\frac{1}{a} e^{a(\zeta - \tau)} = \int^u_{\infty} e^{-a\eta} d\eta,
\label{EQ07}
\end{align}
\begin{align}
\tilde{v} = t + z = & \frac{1}{a} e^{a(\zeta + \tau)} = \int^v_{-\infty} e^{a\eta} d\eta,
\label{EQ08}
\end{align}
where $u = \tau - \zeta$ and $v = \tau + \zeta$. \\
Consider now a massless scalar field. Since the action $\texttt{S}$ is conformally invariant, the action looks similar in both, the inertial and the accelerated frames,
\begin{align}
\texttt{S} & = \frac{1}{2} \int \big( (\partial_{t} \varphi)^2 - (\partial_{z} \varphi)^2  \big) dt dz, \nonumber \\
& = \frac{1}{2} \int \big( (\partial_{\tau} \varphi)^2 - (\partial_{\zeta} \varphi)^2 \big) d\tau d\zeta.
\label{EQ09}
\end{align}
This action defines a Klein-Gordon equation for each spacetime, Minkowski and Rindler, whose solutions are:
\begin{align}
\hat{\varphi}^M = \frac{1}{\sqrt{4\pi \tilde{\omega}}} e^{-i\tilde{\omega}(t - z)},
\label{EQ10}
\end{align}
\begin{align}
\hat{\varphi}^R = \frac{1}{\sqrt{4\pi \omega}} e^{-i\omega(\tau - \zeta)}.
\label{EQ11}
\end{align}
The scalar field is quantized by imposing commutation relations and by rewriting the field in terms of creation and annihilation operators. Both coordinate frames overlap in the domain $z > t$. Therefore, within this domain, the mode expansions for the field operator $\hat{\varphi}$ is (\cite{MVWS}):
\begin{widetext}
\begin{align}
\hat{\varphi} = & \frac{1}{\sqrt{2\pi}}\int \frac{d\tilde{\omega}}{\sqrt{2\tilde{\omega}}} \Big( e^{-i\tilde{\omega}(t - z)}\hat{a}_{\tilde{\omega}} + e^{i\tilde{\omega}(t - z)}\hat{a}^\dagger_{\tilde{\omega}} + e^{-i\tilde{\omega}(t + z)}\hat{a}_{-\tilde{\omega}} + e^{i\tilde{\omega}(t + z)}\hat{a}^\dagger_{-\tilde{\omega}} \Big), \nonumber \\
= & \frac{1}{\sqrt{2\pi}}\int \frac{d\omega}{\sqrt{2\omega}} \Big( e^{-i\omega(\tau - \zeta)}\hat{b}_{\omega} + e^{i\omega(\tau - \zeta)}\hat{b}^\dagger_{\omega} + e^{-i\omega(\tau + \zeta)}\hat{b}_{-\omega} + e^{i\omega(\tau + \zeta)}\hat{b}^\dagger_{-\omega} \Big).
\label{EQ12}
\end{align}
\end{widetext}
The operators $\hat{a}_{\tilde{\omega}}$ and $\hat{b}_{\omega}$ obey the commutation relations:
\begin{align}
[\hat{a}_{\tilde{\omega}}, \hat{a}^\dagger_{\tilde{\omega}'}] = \ \delta(\tilde{\omega} - \tilde{\omega}'), \ \ [\hat{a}_{\tilde{\omega}}, \hat{a}_{\tilde{\omega}'}] = \ 0, \ \ [\hat{a}^\dagger_{\tilde{\omega}}, \hat{a}^\dagger_{\tilde{\omega}'}] = \ 0,
\label{EQ13}
\end{align}
\begin{align}
[\hat{b}_{\omega}, \hat{b}^\dagger_{\omega'}] = \ \delta(\omega - \omega'), \ \ [\hat{b}_{\omega}, \hat{b}_{\omega'}] = \ 0, \ \ [\hat{b}^\dagger_{\omega}, \hat{b}^\dagger_{\omega'}] = \ 0.
\label{EQ14}
\end{align}
The Minkowski $|{0}_{M}\rangle$ and Rindler $|{0}_{R}\rangle$ vacua are defined respectively as $\hat{a}_{\tilde{\omega}}|{0}_{M}\rangle = 0$ and $\hat{b}_{\omega}|{0}_{R}\rangle = 0$.
The creation and annihilation operators are related through the Bogolyubov transformation \cite{BNDP}:
\begin{align}
\hat{b}_{\omega} = \int_{-\infty}^{\infty} d\tilde{\omega} (\alpha_{\tilde{\omega} \omega}\hat{a}_{\tilde{\omega}} - \beta_{\tilde{\omega} \omega}\hat{a}^\dagger_{\tilde{\omega}}).
\label{EQ15}
\end{align}
The normalization condition for the Bogolyubov coefficients is:
\begin{align}
\int_{-\infty}^{\infty} d\tilde{\omega} (\alpha_{\tilde{\omega} \omega}\alpha^\ast_{\tilde{\omega}' \omega} - \beta_{\tilde{\omega} \omega}\beta^\ast_{\tilde{\omega}' \omega}) = \delta(\omega - \omega').
\label{EQ16}
\end{align}
By substituting Eq. (\ref{EQ15}) into Eq. (\ref{EQ12}) and considering
\begin{align}
\int_{-\infty}^{\infty} e^{i(\omega - \omega')u} du = 2\pi\delta(\omega - \omega'),
\label{EQ17}
\end{align}
one can expand the $b$-operators in terms of the $a$-operators
\begin{widetext}
\begin{align}
\hat{b}_{\omega} = \frac{1}{{2\pi}} \int_0^\infty d\tilde{\omega} \sqrt{\frac{\omega}{\tilde{\omega}}}\int_{-\infty}^\infty du \Big( e^{-i\tilde{\omega} \tilde{u} + i\omega u}\hat{a}_{\tilde{\omega}} + e^{i\tilde{\omega} \tilde{u} + i\omega u}\hat{a}^\dagger_{\tilde{\omega}} + e^{-i\tilde{\omega} \tilde{v} + i\omega u}\hat{a}_{-\tilde{\omega}} + e^{i\tilde{\omega} \tilde{v} + i\omega u}\hat{a}^\dagger_{-\tilde{\omega}} \Big).
\label{EQ18}
\end{align}
\end{widetext}
Hence, it follows that the Bogolyoubov coefficients are:
\begin{align}
\alpha_{\tilde{\omega} \omega} = & \frac{1}{2\pi} \sqrt{\frac{\omega}{\tilde{\omega}}}\int_{-\infty}^\infty e^{i(-{\tilde{\omega} \tilde{u}} + {\omega u})} du,
\label{EQ19}
\end{align}
\begin{align}
\alpha_{\tilde{\omega} -\omega} = & \frac{1}{2\pi} \sqrt{\frac{\omega}{\tilde{\omega}}}\int_{-\infty}^\infty e^{i(-{\tilde{\omega} \tilde{v}} + {\omega u})} du,
\label{EQ20}
\end{align}
\begin{align}
\beta_{\tilde{\omega} \omega} = & -\frac{1}{2\pi} \sqrt{\frac{\omega}{\tilde{\omega}}}\int_{-\infty}^\infty e^{i({\tilde{\omega} \tilde{u}} + {\omega u})} du,
\label{EQ21}
\end{align}
\begin{align}
\beta_{\tilde{\omega} -\omega} = & -\frac{1}{2\pi} \sqrt{\frac{\omega}{\tilde{\omega}}}\int_{-\infty}^\infty e^{i({\tilde{\omega} \tilde{v}} + {\omega u})} du.
\label{EQ22}
\end{align}
For the observer moving with uniform acceleration we have $\tilde{u} = -\frac{1}{a} e^{-a u}$ and $\tilde{v} = \frac{1}{a} e^{-a u}$. One can show that $\alpha_{\tilde{\omega} \omega} = -\beta_{\tilde{\omega} -\omega}$ and $\alpha_{\tilde{\omega} -\omega} = -\beta_{\tilde{\omega} \omega}$. Next, by performing the integration in Eqs. (\ref{EQ19})-(\ref{EQ22}), we obtain:
\begin{align}
\alpha_{\tilde{\omega} \omega} = & -e^{\pi\omega/a}\beta_{\tilde \omega \omega},
\label{EQ23}
\end{align}
\begin{align}
\beta_{\tilde{\omega} \omega} = & -\frac{1}{2\pi a} \sqrt{\frac{\omega}{\tilde{\omega}}}\,e^{-\pi\omega/2a} \Big(\frac{\tilde{\omega}}{a} \Big )^{-i\omega/a} \Gamma\Big (\frac{i\omega}{a} \Big).
\label{EQ24}
\end{align}
The vacua $|{0}_{M}\rangle$ and  $|{0}_{R}\rangle$ are different. The Minkowski $a$-vacuum is a state with Rindler $b$-particles and viceversa. The expectation value of the $b$-particle number operator, $\hat{N}_{\omega}$, in the Minkowski vacuum  $|{0}_{M}\rangle$ is:
\begin{align}
\langle{0}_{M}|\hat{N}_{\omega}|{0}_{M}\rangle = \langle{0}_{M}|\hat{b}^{\dagger}_{\omega} \hat{b}_{\omega'}|{0}_{M}\rangle = \int_{-\infty}^{\infty} d\tilde{\omega} |\beta_{\tilde{\omega} \omega}|^2.
\label{EQ25}
\end{align}
This is interpreted as the mean number of particles with frequency $\omega$ found by the accelerated observer. Using Eq. (\ref{EQ21}), with the help of Eq. (\ref{EQ17}), we obtain:
\begin{align}
\int_{-\infty}^{\infty} d\tilde{\omega} |\beta_{\tilde{\omega} \omega}|^2 = \frac{1}{e^{2\pi\omega/a} - 1} \delta(\omega - \omega')\Bigr|_{\substack{\omega' = \omega}}.
\label{EQ26}
\end{align}
This yields the mean number of particles as:
\begin{align}
\langle{0}_{M}|\hat{N}_{\omega}|{0}_{M}\rangle = \frac{\delta(0)}{e^{2\pi\omega/a} - 1}.
\label{EQ27}
\end{align}
The divergent factor, $\delta(0)$, arises due to the infinite volume of the entire space. If the field were quantized within a finite box of volume $V$, the mean density of the particles with frequency $\omega$ would be (\cite{MVWS}):
\begin{align}
n_{\omega} = \frac{\langle{0}_{M}|\hat{N}_{\omega}|{0}_{M}\rangle}{V} = \frac{1}{e^{2\pi\omega/a} - 1}.
\label{EQ28}
\end{align}
%%%%%%%%%%%%%%%%%%%%%%%%%%%%%%%%%%%%%%%%%%%%%%%%%%%%%%%%%%%%%%%%%%%%%%%
\subsection{Non-uniform Rindler spacetime: non-uniformly accelerated observer}
Consider the transformation from the Minkowski spacetime to the non-uniform Rindler spacetime given by:
\begin{align}
t - z = \ \tilde{u} = \int^{\tau} e^{-\chi(\eta)} d\eta - \zeta e^{-\chi(\tau)},
\label{EQ29}
\end{align}
\begin{align}
t + z = \ \tilde{v} = \int^{\tau} e^{\chi(\eta)} d\eta + \zeta e^{\chi(\tau)},
\label{EQ30}
\end{align}
where $\chi(\tau)$ is an arbitrary monotonic function, previously defined as the observer's speed. From here it follows that
\begin{align}
ds^2 = e^{\chi(\bar{v}) - \chi(\bar{u})} (d\tau^2 - d\zeta^2).
\label{EQ31}
\end{align}
Since the metric is conformally invariant, the quantization of the scalar field for the arbitrarily moving observer proceeds analogously to the uniformly accelerated case. The solution of the Klein-Gordon equation defined through Eq. (\ref{EQ06}) for the non-uniform Rindler spacetime is (see \cite{MDAPHDTHESIS_2022}):
\begin{align}
\hat{\varphi}^N = \frac{1}{\sqrt{4\pi\Omega}} e^{-i\Omega \big( \int \cosh{\chi(\tau)} d\tau - \zeta \sinh{\chi(\tau)} \big)},
\label{EQ32}
\end{align}
while the Minkowski one is $\hat{\varphi}^M = \frac{1}{\sqrt{4\pi\omega}} e^{-i\omega(t - z)}$. Both coordinate systems, $(t, z)$ and $(\tau, \zeta)$, are related through Eqs. (\ref{EQ04}) and (\ref{EQ05}).
In the domain of the spacetime where both coordinate frames overlap, one can write the mode expansions for the field operator $\hat{\varphi}$ as follows:
\begin{widetext}
\begin{align}
\hat{\varphi} = & \frac{1}{\sqrt{2\pi}}\int_{0}^{\infty} \frac{d\omega}{\sqrt{2\omega}} \Big( e^{-i\omega(t - z)}\hat{a}_{\omega} + e^{i\omega(t - z)}\hat{a}^\dagger_{\omega} + e^{-i\omega(t + z)}\hat{a}_{-\omega} + e^{i\omega(t + z)}\hat{a}^\dagger_{-\omega} \Big), \nonumber \\
= & \frac{1}{\sqrt{2\pi}}\int_{0}^{\infty} \frac{d\Omega}{\sqrt{2\Omega}} \Big( e^{-i\Omega \big( \int \cosh{\chi(\tau)}d\tau - \zeta\sinh{\chi(\tau)} \big)}\hat{b}_{\Omega} + e^{i\Omega \big( \int \cosh{\chi(\tau)}d\tau - \zeta\sinh{\chi(\tau)} \big)}\hat{b}^\dagger_{\Omega} \ + \nonumber \\
& e^{-i\Omega \big( \int \cosh{\chi(\tau)}d\tau + \zeta\sinh{\chi(\tau)} \big)}\hat{b}_{-\Omega} + e^{i\Omega \big( \int \cosh{\chi(\tau)}d\tau + \zeta\sinh{\chi(\tau)} \big)}\hat{b}^\dagger_{-\Omega} \Big).
\label{EQ33}
\end{align}
\end{widetext}
The lightcone coordinates in Minkowski spacetime are:
\begin{align}
\label{EQ34}
\tilde{U} = t - z, \\
\tilde{V} = t + z.
\label{EQ35}
\end{align}
The lightcone coordinates in the non-uniform Rindler spacetime are:
\begin{align}
\label{EQ36}
U = \int \sinh{\chi(\tau)}d\tau - \zeta\cosh{\chi(\tau)}, \\
V = \int \cosh{\chi(\tau)}d\tau - \zeta\sinh{\chi(\tau)}.
\label{EQ37}
\end{align}
If we look at the hypersurface located at the lightcone $t = z$, the lightcone coordinates $U$ and $V$ give rise to a Rindler mode $\sim e^{i \omega (U + V)}$ which is a combination of both lightcone normal modes. Through the surface where $\tilde{V} = \infty$ and $\tilde{U} = 0$ we obtain a mode which is proportional to the previous mode. Considering this, now we find that (see \cite{UWG}, \cite{BNDP}, \cite{FSA}, \cite{BQFT1975}, \cite{CLH}):
\begin{widetext}
\begin{align}
\frac{1}{\sqrt{\Omega}} e^{i\Omega\int \cosh{\chi(\tau)}d\tau + i\Omega\zeta\sinh{\chi(\tau)}} = \int \frac{d\omega}{\sqrt{\omega}} \big( \alpha_{\omega \Omega} e^{-i\omega (\zeta + e^{a\zeta}/{a}) + i\omega (\tau - \zeta\sinh{a\tau})} - \beta_{\omega \Omega} e^{i\omega (\zeta + e^{a\zeta}/{a}) - i\omega (\tau - \zeta\cosh{a\tau})} \big).
\label{EQ38}
\end{align}
\end{widetext}
Multiplying both sides of this equation by $e^{i\Omega\zeta - i\omega' (\zeta + e^{a\zeta}/{a})}$ and integrating over $\zeta$ we obtain:
\begin{widetext}
\begin{align}
e^{i\Omega\int d\tau\cosh{\chi(\tau)}} \int d\zeta \big( e^{i\Omega\zeta - {i\omega' e^{a\zeta}}/{a}} \big) \big( e^{i\Omega\zeta\sinh{\chi(\tau)} - i\omega'\zeta} \big) = -\int d\omega e^{-i\omega\tau} \sqrt{\frac{\Omega}{\omega}} \beta_{\omega \Omega} \Big( \int d\zeta e^{i(\zeta + e^{a\zeta}/{a})(\omega - \omega')} e^{i(\Omega + \omega\cosh{a\tau}) \zeta} \Big).
\label{EQ39}
\end{align}
\end{widetext}
Hence, a straightforward calculation of the integrals, yields the Bogolyubov coefficient as:
\begin{widetext}
\begin{align}
\beta_{\omega \Omega} = -\sqrt{\frac{\Omega}{|a + a\cosh{a\tau}|}} \sqrt{\frac{\Omega}{\omega}} e^{-\pi \Omega/2a} \Gamma\Big(\frac{i\Omega}{a}\Big) \Big(\frac{\omega}{a}\Big)^{-i\Omega/a} \delta \big( \omega - \Omega \sinh{\chi(\tau)} \big).
\label{EQ40}
\end{align}
\end{widetext}
In order to find the mean number of particles, $\mathcal{N}_{\Omega} = \langle{0}_{M}|\hat{N}_{\Omega}|{0}_{M}\rangle$, we trace out the Minkowski momenta:
\begin{align}
\mathcal{N}_{\Omega} = \int_{-\infty}^{\infty} d\omega |\beta_{\omega \Omega}|^2.
\label{EQ41}
\end{align}
After some algebra, we obtain:
\begin{align}
\int_{-\infty}^{\infty} d\omega |\beta_{\omega \Omega}|^2 = \frac{\sinh{\chi(\tau)}}{|1 + \cosh{a\tau}|} \bigg( \frac{\delta(0)}{e^{2\pi\Omega/a} - 1} \bigg).
\label{EQ42}
\end{align}
Accordingly,
\begin{align}
\mathcal{N}_{\Omega} = \frac{\sinh{\chi(\tau)}}{|1 + \cosh{a\tau}|} \bigg( \frac{\delta(0)}{e^{2\pi\Omega/a} - 1} \bigg).
\label{EQ43}
\end{align}
Thus, considering again that $V = \delta(0)$, the mean density of particles is given by $\mathbcal{n}_{\Omega} = \mathcal{N}_{\Omega}/V$. Hence:
\begin{align}
\mathbcal{n}_{\Omega} = \frac{\sinh{\chi(\tau)}}{|1 + \cosh{a\tau}|} \bigg( \frac{1}{e^{2\pi\Omega/a} - 1} \bigg).
\label{EQ44}
\end{align}
%%%%%%%%%%%%%%%%%%%%%%%%%%%%%%%%%%%%%%%%%%%%%%%%%%%%%%%%%%%%%%%%%%%%%%%
%%%%%%%%%%%%%%%%%%%%%%%%%%%%%%%%%%%%%%%%%%%%%%%%%%%%%%%%%%%%%%%%%%%%%%%
%%%%%%%%%%%%%%%%%%%%%%%%%%%%%%%%%%%%%%%%%%%%%%%%%%%%%%%%%%%%%%%%%%%%%%%
\section{The density of particles for the non-uniform Rindler spacetime}

Equation (\ref{EQ44}) provides the particle density observed by a detector undergoing non-uniform acceleration. To verify this result, we consider limiting cases with well-established particle density behavior. To this end, we define a specific form for $\chi(\tau)$ as follows (see the green lines in Fig. \ref{Fig1}):
\begin{align}
\chi(\tau) = \ln (1 + e^{a \tau}).
\label{EQ45}
\end{align}
 For this $\chi(\tau)$ we obtain the acceleration
\begin{align}
\alpha(\tau) = \frac{a}{2}\bigg(1 + \tanh \Big( \frac{a\tau}{2} \Big)\bigg).
\label{EQ46}
\end{align}
The density of particles becomes now:
\begin{align}
\mathbcal{n}_{\Omega} = \frac{\sinh{\big( \ln (1 + e^{a \tau}) \big)}}{|1 + \cosh{a\tau}|} \bigg( \frac{1}{e^{2\pi \Omega/a} - 1} \bigg).
\label{EQ47}
\end{align}
This expression shows two special features to the observer's reference frame (non-uniform Rindler spacetime): i) for $\tau \to -\infty$ the observer proper spacetime matches exactly the Minkowski spacetime, and ii) for $\tau \to +\infty$ the observer proper spacetime coincides exactly with the Rindler spacetime.
%%%%%%%%%%%%%%%%%%%%%%%%%%%%%%%%%%%%%%%%%%%%%%%%%%%%%%%%%%%%%%%%%%%%%%%
\subsection{Density of particles for $\tau \to -\infty$}
In this case, we have $U(\tau \to -\infty) = \tau$ and $\tilde{U}(\tau \to -\infty) = \tau - \zeta$. Then, we obtain:
\begin{align}
e^{i\Omega\tau} = \int d\omega \sqrt{\frac{\Omega}{\omega}} \Big( \alpha_{\omega \Omega} e^{i\omega(\tau - \zeta)} + \beta_{\omega \Omega} e^{-i\omega(\tau - \zeta)} \Big).
\label{EQ48}
\end{align}
Multiplying both sides of this equation by $e^{-i\omega\tau}$ and integrating over $\tau$ and then multiplying both sides of this equation by $e^{-i\omega\zeta}$ and integrating over $\zeta$ we obtain:
\begin{align}
\beta_{\omega \Omega} = -\delta(\omega).
\label{EQ49}
\end{align}
This result leads us to:
\begin{align}
\mathbcal{n}_{\Omega} = \frac{1}{V} \int_{-\infty}^{\infty} d\omega |\beta_{\Omega \omega}|^2 = 0,
\label{EQ50}
\end{align}
which matches with the corresponding result from Eq. (\ref{EQ47}):
\begin{align}
\lim_{\tau \rightarrow -\infty} \frac{\sinh{\big( \ln (1 + e^{a \tau}) \big)}}{|1 + \cosh{a\tau}|} = 0, \nonumber
\label{}
\end{align}
hence $\mathbcal{n}_{\Omega} = 0$. This result is anticipated because, in the limit $\tau \to -\infty$, the non-uniform Rindler spacetime converges to the Minkowski spacetime, where the vacuum is uniquely defined and devoid of particles for all inertial observers.
%%%%%%%%%%%%%%%%%%%%%%%%%%%%%%%%%%%%%%%%%%%%%%%%%%%%%%%%%%%%%%%%%%%%%%%
\subsection{Density of particles for $\tau \to +\infty$}
In this case we must workout a slightly different procedure, considering that in this limit both sides of Eq. (\ref{EQ38}) exhibit a non-trivial relationship. To circumvent this complexity, we employ the following approach. If we look at the hypersurface located in the lightcone $t = z$, the lightcone coordinates, $U$ and $V$, give rise to a Rindler mode $\sim e^{i\omega(U + V)}$ which is a combination of both lightcone normal modes. Through the surface where $\tilde{V} = \infty$ and $\tilde{U} = 0$ we obtain a mode which is proportional to the previous mode. Considering this, now we use
\begin{align}
\frac{1}{\sqrt{\Omega}} e^{i\Omega V} = \int \frac{d\omega}{\sqrt{\omega}} \Big( \alpha_{\omega \Omega} e^{-i\omega\tilde{V}} + \beta_{\omega \Omega} e^{i\omega\tilde{V}} \Big),
\label{EQ51}
\end{align}
where $V = \int \cosh{\chi(\tau)}d\tau - \zeta\sinh{\chi(\tau)}$ and $\tilde{V} = t + z$. Then we have $V(\tau \to +\infty, \zeta = 0) = \tau + {e^{a\tau}}/{2a}$ and $\tilde{V}(\tau \to +\infty, \zeta = 0) = \tau + {e^{a\tau}}/{a}$.
Rearranging terms and multiplying both sides of this equation by $e^{-i\omega\tau}$ and integrating over $\tau$ we obtain:
\begin{align}
& \int d\tau \big( e^{i\Omega\tau - {i\omega e^{a\tau}}/{a}} \big) \big( e^{{i\Omega e^{a\tau}}/{2a} - i\omega\tau} \big) = \nonumber \\
& \int d\omega' \sqrt{\frac{\Omega}{\omega'}} \beta_{\omega' \Omega} \Big( \int d\tau e^{i(\omega' - \omega)\tau} \Big).
\label{EQ52}
\end{align}
Performing an integration by parts in the LHS and the two single integrations in the RHS of this equation we obtain:
\begin{align}
\beta_{\omega \Omega} = \frac{1}{2\pi a} \sqrt{\frac{\Omega}{\omega}} e^{-\pi\Omega/{2a}} \Big( \frac{\omega}{a} \Big) ^{-i\Omega/a} \Gamma\Big( \frac{i\Omega}{a} \Big) F_{div},
\label{EQ53}
\end{align}
where $F_{div}$ is a divergent factor which has an asymptotic behavior: $F_{div} \sim V^{1/2}$. Performing an integration by parts in the mean density of particles
\begin{align}
\mathbcal{n}_{\Omega} = \frac{1}{V} \int_{-\infty}^{\infty} d\omega |\beta_{\Omega \omega}|^2,
\label{EQ54}
\end{align}
the calculations leads us to:
\begin{align}
\mathbcal{n}_{\Omega} = \bigg( \frac{1}{e^{2\pi\Omega/a} - 1} \bigg).
\label{EQ55}
\end{align}
This result clearly matches the well-established result for the Rindler spacetime. Corroborating the corresponding result from Eq. (\ref{EQ47}):
\begin{align}
\lim_{\tau \rightarrow +\infty} \frac{\sinh{\big( \ln (1 + e^{a \tau}) \big)}}{|1 + \cosh{a\tau}|} = 1, \nonumber
\label{}
\end{align}
hence obtaining for Eq. (\ref{EQ47}) the same expression as the corresponding one for Eq. (\ref{EQ55}).
%%%%%%%%%%%%%%%%%%%%%%%%%%%%%%%%%%%%%%%%%%%%%%%%%%%%%%%%%%%%%%%%%%%%%%%
%%%%%%%%%%%%%%%%%%%%%%%%%%%%%%%%%%%%%%%%%%%%%%%%%%%%%%%%%%%%%%%%%%%%%%%
%%%%%%%%%%%%%%%%%%%%%%%%%%%%%%%%%%%%%%%%%%%%%%%%%%%%%%%%%%%%%%%%%%%%%%%
\section{Quantum entanglement and the Rindler spacetime}

There are certain quantum states $|\Psi\rangle$ of composite systems in which the state can be expressed as the tensor product of the states of the individual elements. For the initial system detector+field, the state can be written as (\cite{PhysRevD.96.083531}, \cite{2016MSRR}, \cite{2014MM}):
\begin{align}
|\Psi_0\rangle = |\psi\rangle \otimes |0_{\texttt{SB}}\rangle,
\label{EQ56}
\end{align}
where $|\psi\rangle$ is the initial state of the detector and $|0_{\texttt{SB}}\rangle$ is the inertial spacetime background vacuum for the field. This state $|\Psi_{0}\rangle$ is a \textit{pure state}  to the extent that it can be represented as the tensor product of the individual states of the originally separated elements. However, other states cannot be expressed in this way under any circumstance. Those states of composite systems that cannot be written as the tensor products of the individual states are known as \textit{entangled states} (\cite{PhysRevD.96.083531}).

This entanglement of states becomes evident when measurements on a composite system yield well-defined global outcomes, while individual properties remain ill-defined. Measuring one element of the entangled system determines (through a global correspondence) the corresponding outcome of the other, regardless of spatial separation. This correlation, while seemingly enabling faster-than-light communication, is bound by the uncertainty principle. Measuring one element does not collapse the wave function or affect the uncertainty of the other until both measurements are completed (\cite{PhysRevD.96.083531}, \cite{2016MSRR}, \cite{2014MM}).

Let us restrict ourselves to systems of two elements $A+B$ (bipartite systems), mirroring the system detector+field of the previous chapter. Let $\rho_{\texttt{AB}} = |\Psi\rangle \langle\Psi|$ be the density matrix of the entire system. When we trace out one of the elements (B) of the bipartite system,
\begin{align}
\rho_{\texttt{A}} = Tr_{\texttt{B}}\rho_{\texttt{AB}},
\label{EQ57}
\end{align}
the other element (A) shows a non-pure state; given that the state of the whole system is a pure state, the mixing must be generated by the entanglement of the elements of the system. The simplest way to quantify the entanglement of a bipartite systems is through the Von Neumann entropy. The Von Neumann entropy of the reduced density matrix serves as a measure of the entanglement between two subsystems within a composite quantum system. It quantifies the level of entanglement by computing the entropy of the reduced density matrix for each subsystem. A non-zero entropy indicates the presence of entanglement between the subsystems. For bipartite pure quantum states, the Von Neumann entropy of the reduced states is the unique measure of entanglement that satisfies certain axioms required for an entanglement measure (\cite{PhysRevD.96.083531}, \cite{2016MSRR}, \cite{2014MM}). Hence, we define the Von Neumann entropy of one subsystem as:
\begin{align}
S_{\texttt{VN}}(\rho_A) = -\rm{Tr} (\rho_{\texttt{A}} \ln \rho_{\texttt{A}}),
\label{EQ58}
\end{align}
or, if we know the eigenvalues $\lambda_i^A$ of $\rho_{\texttt{A}}$, the latter expression can be written as:
\begin{align}
S_{\texttt{VN}}(\rho_{\texttt{A}}) = -\sum_i \lambda_i^A \ln \lambda_i^A.
\label{EQ59}
\end{align}
The mathematical structure of this expression reveals that entanglement increases with greater mixing and correlation between elements. The traceness of the $B$ element in the bipartite system induces entanglement, and the entanglement can be measured using the Von Neumann entropy of the reduced density matrix (\cite{PhysRevD.96.083531}). It is important to note that there exist other measures of entanglement, such as relative entropy of entanglement, robustness of entanglement, and squashed entanglement, which can be generalized to the multipartite scenario. However, for bipartite pure states, the Von Neumann entropy of the reduced states remains the unique measure for bipartite pure states.

The Unruh effect, however, is not the only quantum consequence arising from non-inertial observations of the Minkowski vacuum. Quantum entanglement within the vacuum itself presents a significant consequence. While different inertial observers will perceive different values for the reduced Von Neumann entropy $S_{vn}(\rho_{\texttt{A}})$, and entanglement entropy of a subsystem $A$, the total entanglement entropy remains invariant (\cite{PhysRevLett.88.230402, PhysRevLett.89.270402}). Nevertheless, this invariance does not hold for non-inertial observers. It has been suggested (\cite{PhysRevLett.95.120404}) that the Unruh effect diminishes the entanglement across the event horizon, highlighting the observer-dependent nature of entanglement in non-inertial frames. \\
%%%%%%%%%%%%%%%%%%%%%%%%%%%%%%%%%%%%%%%%%%%%%%%%%%%%%%%%%%%%%%%%%%%%%%%
\subsection{Entanglement generation via the Unruh effect in Rindler spacetime}

\begin{figure}
\begin{center}
\scalebox{0.40}{\includegraphics{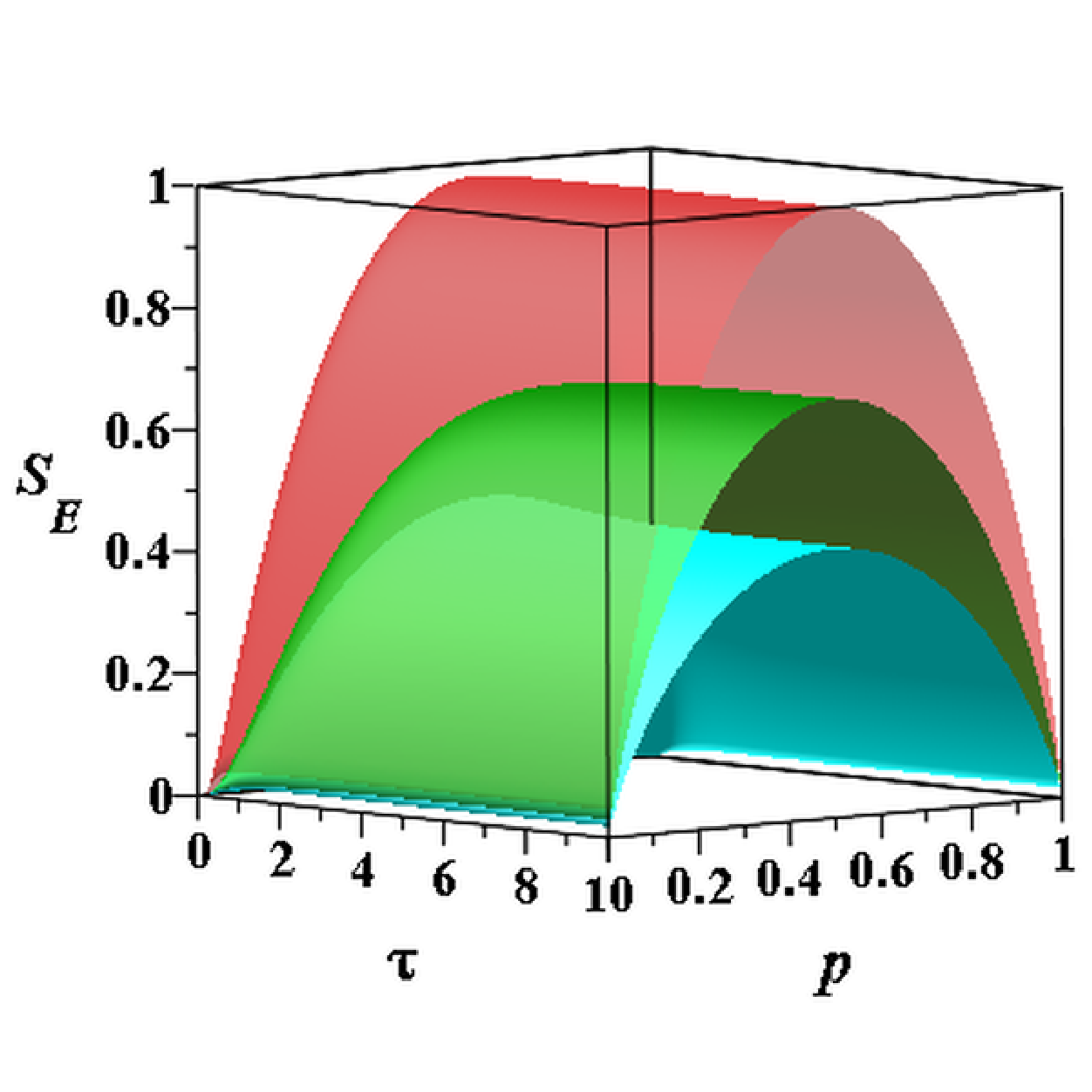}}
\end{center}
\caption{The entanglement entropy, $S_E$, as a function of $\tau$ and $p$. The red surface corresponds to $\eta = 10$.  The green surface corresponds to $\eta = 1$. The cyan surface corresponds to $\eta = 0$.
\label{Fig2}}
\end{figure}

How quantum-mechanical non-locality depends on the global structure of spacetime? From the point of view of local observers, the vacuum state of any quantum field is entangled, and thus localized vacuum fluctuations are correlated (\cite{PATD, RevModPhys.73.565}). For bipartite system, ${Sys}_{\texttt{AB}} = {Sys}_{\texttt{A}} \cup {Sys}_{\texttt{B}}$, the entanglement entropy is defined as the von Neumann entropy of the reduced density matrix of either of its parts (\cite{Bennett}). Let $\rho_{\texttt{AB}} = |\Psi\rangle \langle\Psi|$ be again the density matrix of the entire system. Then, the entanglement entropy of the subsystem ${Sys}_{\texttt{A}}$ is given by:
\begin{align}
S_{\texttt{E}}(\rho_{\texttt{A}}) = - Tr (\rho_{\texttt{A}} \ln \rho_{\texttt{A}}).
\label{EQ60}
\end{align}
In a similar way we can define $S_{\texttt{E}}(\rho_{\texttt{B}})$, and it can be show that $S_{\texttt{E}}(\rho_{\texttt{A}}) = S_{\texttt{E}}(\rho_{\texttt{B}}) = S_{\texttt{E}}$ (\cite{MDAPHDTHESIS_2022}). We state that the density matrix (in the interaction picture) is (see \cite{AINGPBMDAXW_DADOUDWF_2020, MDAPHDTHESIS_2022}):
\begin{align}
\label{EQ61}
\rho_{ij}(\tau) = & \langle i | {\rm Tr}_{\texttt{F}} U(\tau,\tau_0)\varrho(\tau_0) U^{-1}(\tau,\tau_0)| j \rangle. \\
& (i,j = 0,1 = \downarrow \uparrow), \nonumber
\end{align}
where we trace out the scalar field degrees of freedom over the evolution operator $U(\tau,\tau_0)$ of the system for the initial time $\tau_0$ and the subsequent time $\tau$. Hence:
\begin{align}
\rho(\tau) = \begin{pmatrix}
p & \sqrt{p(1-p)} e^{-\gamma (\tau)} \\ 
\sqrt{p(1-p)} e^{-\gamma (\tau)} & (1-p)
\end{pmatrix},
\label{EQ62}
\end{align}
where $p = |\langle\uparrow|\psi\rangle|^2$ and $\gamma (\tau)$ is the decoherence function derived in our previous work \cite{AINGPBMDAXW_DADOUDWF_2020}. By diagonalizing this matrix, we obtain (\cite{MDAPHDTHESIS_2022}):
\begin{align}
\rho(\tau) = \begin{pmatrix}
\lambda_{1}(\tau) & 0 \\ 
0 & \lambda_{2}(\tau)
\end{pmatrix},
\label{EQ63}
\end{align}
whose eigenvalues are
\begin{align}
\lambda_{1, 2}(\tau) = \frac{1 \pm r(\tau)}{2},
\label{EQ64}
\end{align}
where
\begin{align}
r(\tau) = \sqrt{1 - 4p(1-p)(1-e^{-2\gamma (\tau)})}.
\label{EQ65}
\end{align}
The computation of the entanglement entropy of the detector,
\begin{align}
S_{\texttt{E}}(\tau) = -\sum_i \lambda_i \ln \lambda_i,
\label{EQ66}
\end{align}
yields:
\begin{widetext}
\begin{align}
S_{\texttt{E}}(\tau) = 1 - \tfrac12 \big(1+ r(\tau) \big) \ln \big(1+ r(\tau)\big) - \tfrac12 \big(1- r(\tau) \big) \ln \big(1- r(\tau)\big).
\label{EQ67}
\end{align}
\end{widetext}
The results of the numerical simulation are presented in Fig. \ref{Fig2}. The acceleration produces an additional entanglement entropy of the detector due to the interaction with Unruh radiation (green and red surfaces). Large accelerations means high effective temperatures, which leads to the maximum production of entanglement entropy.

\begin{figure}
\begin{center}
\scalebox{0.50}{\includegraphics{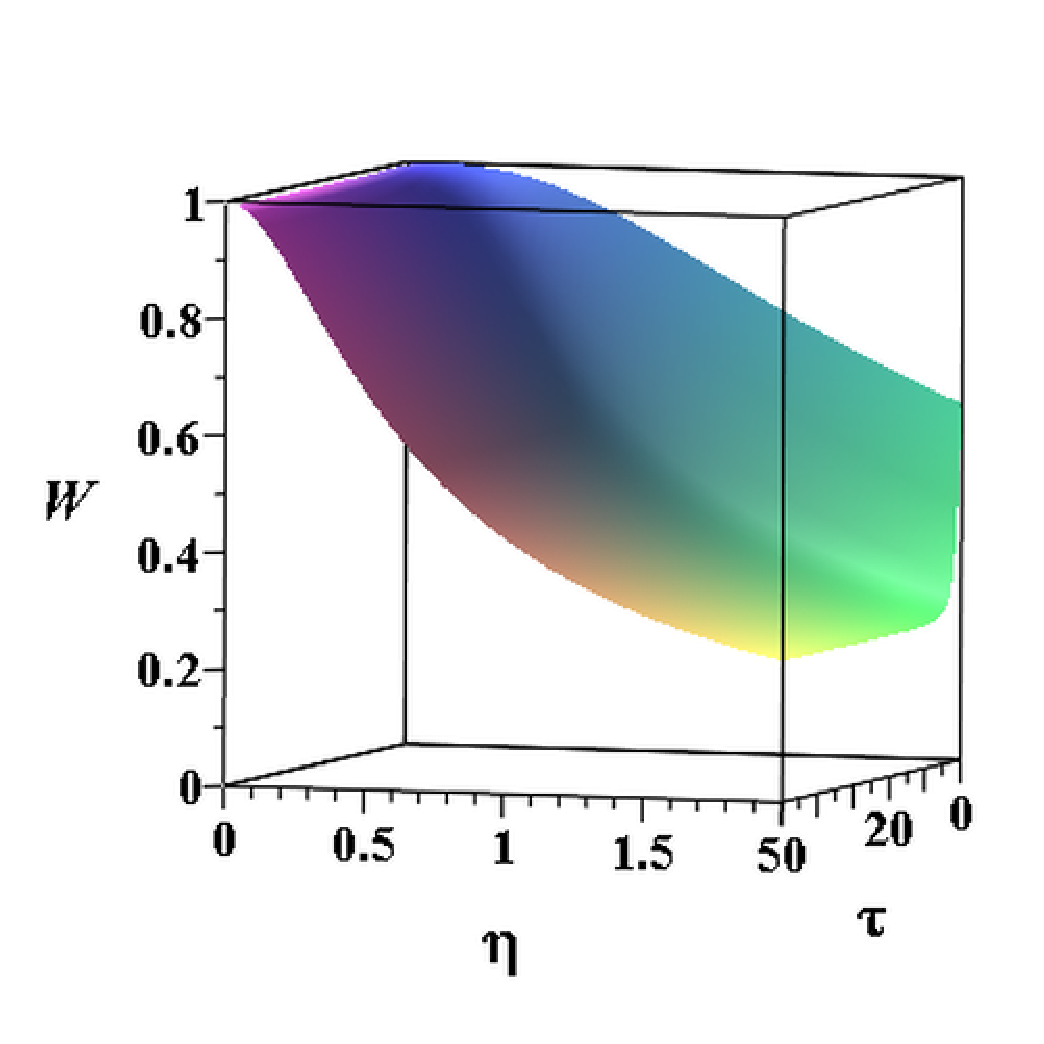}}
\end{center}
\caption{The ratio, $W$, as a function of $\tau$ and $\eta$, indicates that decoherence occurs for the accelerated observer compared to its inertial counterpart.
\label{Fig3}}
\end{figure}

Let's now consider a hypothetical experimental setup in which two identical detectors are coupled to the scalar field, each in both inertial and accelerated reference frames. The experiment involves comparing the partial decoherence for the inertial and the accelerated observers. For this purpose, we introduce a new function, the ratio $W$:
\begin{align}
W = \frac{\ln|\rho^{in}_{01} (\tau)|}{\ln |\rho^{ac}_{01} (\tau)|} \equiv \frac{\gamma(\tau, 0)}{\gamma(\tau, \eta)},
\label{EQ68}
\end{align}
where $\rho^{ac}_{01} (\tau)$ and  $\rho^{in}_{01} (\tau)$ are the components of the density matrix for the accelerated and the inertial observers, respectively, and the function $\gamma(\tau, \eta)$ is defined as (\cite{AINGPBMDAXW_DADOUDWF_2020})
\begin{align}
& \gamma(\tau, \eta) = \frac{2\Lambda_{ms}^2}{\pi^2} \Re \ln (1 + i \tau / l) \nonumber \\
& - \frac{4\Lambda_{ms}^2}{\pi^2} \Re \ln \bigg( \frac{\Gamma (1 + \eta c^2 / 2\pi + i \eta c^2 \tau / (2\pi l))}{\Gamma (1 + \eta c^2 / 2\pi)} \bigg), \nonumber
%\label{EQ}
\end{align}
where $\Gamma(z)$ is the gamma function and $\Lambda_{ms}$ is the scalar field coupling constant, $\eta = al/c^2$, with $a$ as the uniform acceleration of the observer and $l$ as the size of the detector. In Fig. \ref{Fig3}, $W$ is depicted as a function of $\tau$ and $\eta$, where quantum decoherence becomes more pronounced for the accelerated observer compared to the inertial observer.
%%%%%%%%%%%%%%%%%%%%%%%%%%%%%%%%%%%%%%%%%%%%%%%%%%%%%%%%%%%%%%%%%%%%%%%
\subsection{The deformation of the Minkowski vacuum in the Rindler spacetime}
Our aim is to discuss the implications of the Unruh effect on the structure of Rindler spacetime. Firstly, we compute the quantum entanglement of the Minkowski vacuum as perceived by a Rindler observer. We need to establish the relations between the Rindler operators, $\hat{b}_{\Omega}$ and $\hat{b}_{\Omega}^\dagger$, and their Minkowski counterparts, $\hat{a}_{\omega}$ and $\hat{a}_{\omega}^\dagger$; hence (\cite{CLH}):
\begin{align}
\Big( \hat{b}_{\Omega}^{(\pm)} - e^{-\pi \Omega / a} \hat{b}_{\Omega}^{(\mp) \dagger} \Big) | 0_{M} \rangle = 0,
\label{EQ69}
\end{align}
where the signs on the operators denotes the right Rindler wedge $(+)$ and left Rindler wedge $(-)$.
The next step is to expand the Minkowski vacuum in terms of the Rindler eigenstates $| N_{R, n_{k}^{-}, n_{k}^{+}} \rangle = | N_{R, n_{1}^{-}}, N_{R, n_{2}^{-}}, ... \rangle \otimes | N_{R, n_{1}^{+}}, N_{R, n_{2}^{+}}, ... \rangle$. Due to the symmetric distribution of particles on each Rindler wedge, each $n_{k}^{-}$ on the left Rindler wedge must be equal to the corresponding $n_{k}^{+}$ on the right Rindler wedge. Then, we rewrite the Rindler state as
\begin{align}
| N_{R, n_{k}^{-}, n_{k}^{+}} \rangle = | N_{R, n_{k}, n_{k}} \rangle \equiv | N_{R, n_{k}} \rangle.
\end{align}
\label{EQ70}
We state the definition of the Rindler states in terms of its corresponding Rindler vacuum $| 0_{R} \rangle$ as (\cite{MDAPHDTHESIS_2022}):
\begin{align}
| N_{R, n_{k}} \rangle = \bigg( \prod_{k} \frac{(b_{\Omega, k}^{(-) \dagger})^{n_k} (b_{\Omega, k}^{(+) \dagger})^{n_k}}{(n_{k})!} \bigg) | 0_{R} \rangle,
\label{EQ71}
\end{align}
We know (\cite{BNDP}) that the commutation relation for the Rindler annihilation and creation operators is
\begin{align}
[\hat{b}_{\Omega}^{(\pm)}, \hat{b}_{\Omega}^{(\pm) \dagger}]= 1,
\label{EQ72}
\end{align}
which, in combination with Eq. \eqref{EQ69}, yields:
\begin{align}
| N_{R, n_{k}} \rangle & = \bigg( \prod_{k} e^{-n_{k} \pi \Omega_{k} / a} \frac{(b_{\Omega, k}^{(\pm) \dagger})^{n_k} (b_{\Omega, k}^{(\pm) \dagger})^{n_k}}{(n_{k})!} \bigg) | 0_{R} \rangle \nonumber \\
& = \Big( \prod_{k} \sum_{n_{k}} e^{-n_{k} \pi \Omega_{k} / a} \Big) | 0_{R} \rangle.
\label{EQ73}
\end{align}
Expanding the Minkowski vacuum in Rindler eigenstates
\begin{align}
| 0_{M} \rangle = \sum_{n_{k}} | N_{R, n_{k}} \rangle \langle N_{R, n_{k}} | 0_{M} \rangle,
\label{EQ74}
\end{align}
or:
\begin{align}
| 0_{M} \rangle = \langle 0_{R} | 0_{M} \rangle \prod_{k} \Big( \sum_{n_{k}} e^{-n_{k} \pi \Omega_{k} / a} \Big) | N_{R, n_{k}} \rangle.
\label{EQ75}
\end{align}
Equivalently:
\begin{align}
| 0_{M} \rangle = \langle 0_{R} | 0_{M} \rangle \prod_{k} \Big( \sum_{n_{k}} e^{-n_{k} \pi \Omega_{k} / a} \Big) | N_{R, n_{k}^{-}} \rangle \otimes | N_{R, n_{k}^{+}} \rangle.
\label{EQ76}
\end{align}
If we multiply this expression on the left by the Minkowski vacuum state $| 0_{M} \rangle$ and make use of Eq. \eqref{EQ73}, we obtain
\begin{align}
\langle 0_{M} | 0_{M} \rangle = |\langle 0_{R} | 0_{M} \rangle|^2 \prod_{k} \Big( \sum_{n_{k}} e^{-2n_{k} \pi \Omega_{k} / a} \Big) = 1.
\label{EQ77}
\end{align}
It is easy to prove that
\begin{align}
\sum_{n_{k}} \big( e^{-2n_{k} \pi \Omega_{k} / a} \big) = 1 - e^{-2 \pi \Omega_{k} / a},
\label{EQ78}
\end{align}
hence, we define a new quantity $Q_{R}$ as:
\begin{align}
Q_{R} = |\langle 0_{R} | 0_{M} \rangle|^2 = \prod_{k} \big( 1 - e^{-2 \pi \Omega_{k} / a} \big)^{-1},
\label{EQ79}
\end{align}
which, along with Eq. \eqref{EQ73} and Eq. \eqref{EQ76}, yields:
\begin{align}
| 0_{M} \rangle = Q_{R}^{\frac{1}{2}} \prod_{k} \bigg( \sum_{n_{k}} e^{-n_{k} \pi \Omega_{k} / a} \frac{(b_{\Omega, k}^{(\pm) \dagger})^{n_k} (b_{\Omega, k}^{(\mp) \dagger})^{n_k}}{(n_{k})!} \bigg) | 0_{R} \rangle.
\label{EQ80}
\end{align}
It is also easy to prove that
\begin{align}
& \sum_{n_{k}}\bigg( e^{-n_{k} \pi \Omega_{k} / a} \frac{(b_{\Omega, k}^{(\pm) \dagger})^{n_k} (b_{\Omega, k}^{(\mp) \dagger})^{n_k}}{(n_{k})!} \bigg) | 0_{R} \rangle = \nonumber \\
& \exp{\Big( e^{-\pi \Omega_{k} / a} b_{\Omega, k}^{(\pm) \dagger} b_{\Omega, k}^{(\mp) \dagger} \Big)} | 0_{R} \rangle;
\label{EQ81}
\end{align}
then, we express the Minkowski vacuum as:
\begin{align}
| 0_{M} \rangle = Q_{R}^{\frac{1}{2}} \prod_{k} \exp{\Big( e^{-\pi \Omega_{k} / a} b_{\Omega, k}^{(\pm) \dagger} b_{\Omega, k}^{(\mp) \dagger} \Big)} | 0_{R} \rangle,
\label{EQ82}
\end{align}
a two-mode squeezed state. \\
%%%%%%%%%%%%%%%%%%%%%%%%%%%%%%%%%%%%%%%%%%%%%%%%%%%%%%%%%%%%%%%%%%%%%%%
\subsection{The deformation of the Minkowski vacuum in the non-uniform Rindler spacetime}
To illuminate the consequences of the Unruh effect on the structure of non-uniform Rindler spacetime, we must first ascertain the quantum entanglement of the Minkowski vacuum as perceived by a non-inertial observer. This requires establishing the relationship between the non-uniform Rindler annihilation and creation operators that annihilate the Minkowski vacuum. The corresponding Bogolyubov coefficients can be expressed as follows (see \cite{MDAPHDTHESIS_2022}):
\begin{widetext}
\begin{align}
\label{EQ83}
\alpha_{\Omega \omega} & = \sqrt{\frac{\Omega}{a | 1 + \sinh{a\tau} + \sinh{\chi(\tau)} |}} \sqrt{\frac{\Omega}{\omega}} e^{\pi \Omega / 2a} \Gamma \Big(-\frac{i\Omega}{a}\Big) \Big(\frac{\omega}{a}\Big)^{i\Omega/a} \delta \big( \omega - \Omega \sinh{\chi(\tau)} \big) \nonumber \\
& = \sqrt{\frac{\Omega}{a | 1 + \sinh{a\tau} + \sinh{\chi(\tau)} |}} \ e^{\pi \Omega / 2a} F^{\ast}(\omega, \Omega, \tau), \\
\beta_{\Omega \omega} & = -\sqrt{\frac{\Omega}{a | 1 + \cosh{a\tau} |}} \sqrt{\frac{\Omega}{\omega}} e^{-\pi \Omega / 2a} \Gamma \Big(\frac{i\Omega}{a}\Big) \Big(\frac{\omega}{a}\Big)^{-i\Omega/a} \delta \big( \omega - \Omega \sinh{\chi(\tau)} \big) \nonumber \\
& = -\sqrt{\frac{\Omega}{a | 1 + \cosh{a\tau} |}} \ e^{-\pi \Omega / 2a} F(\omega, \Omega, \tau).
\label{E84}
\end{align}
\end{widetext}
The Bogolyubov relations for the annihilation and creation operators in both frameworks are (\cite{BNDP}):
\begin{align}
\hat{c}_{\Omega} = \int d\omega ( \alpha_{\Omega \omega} \hat{a}_{\omega} - \beta_{\Omega \omega} \hat{a}_{\omega}^\dagger ),
\label{EQ85}
\end{align}
and
\begin{align}
\hat{c}_{\Omega}^\dagger = \int d\omega ( \alpha_{\Omega \omega}^\ast \hat{a}_{\omega}^\dagger - \beta_{\Omega \omega}^\ast \hat{a}_{\omega} ).
\label{EQ86}
\end{align}
From here, we can eliminate the terms in $\hat{a}_{\omega}^\dagger$ and find:
\begin{widetext}
\begin{align}
\hat{c}_{\Omega}^\dagger - e^{\pi \Omega / a} \sqrt{\frac{|1 + \sinh{a\tau} + \sinh{\chi(\tau)}|}{|1 + \cosh{a\tau}|}} \ \hat{c}_{\Omega} = e^{\pi \Omega / 2a} \Big( \frac{e^{-\pi \Omega / a}}{|1 + \cosh{a\tau}|} - \frac{\sqrt{|1 + \cosh{a\tau}|}}{|1 + \sinh{a\tau} + \sinh{\chi(\tau)}|} e^{\pi \Omega / a} \Big) \Big( \int d\omega F(\omega, \Omega, \tau) \Big) \hat{a}_{\omega}.
\label{EQ87}
\end{align}
\end{widetext}
Then, we obtain:
\begin{align}
& \bigg( \hat{c}_{\Omega} - e^{-\pi \Omega / a} \sqrt{\frac{|1 + \sinh{a\tau} + \sinh{\chi(\tau)}|}{|1 + \cosh{a\tau}|}} \ \hat{c}_{\Omega}^\dagger \bigg) | 0_{M} \rangle = \nonumber \\
& \Big( \hat{c}_{\Omega} - e^{-\pi \Omega / a} G(\tau) \hat{c}_{\Omega}^\dagger \Big) | 0_{M} \rangle = 0,
\label{EQ88}
\end{align}
where $G(\tau) = \sqrt{\frac{|1 + \sinh{a\tau} + \sinh{\chi(\tau)}|}{|1 + \cosh{a\tau}|}}$. \\
Now, we expand the Minkowski vacuum in terms of the non-uniform Rindler eigenstates $| N_{nuR, n_{k}} \rangle = | N_{nuR, n_1}, N_{nuR, n_2}, ... \rangle$. To begin with, we ascertain that
\begin{align}
| N_{nuR, n_{k}} \rangle = \frac{(c_{\Omega, 1}^\dagger)^{n_1} (c_{\Omega, 2}^\dagger)^{n_2} ...}{(n_{1}! \ n_{2}! \ ...)^{1/2}} | 0_{nuR} \rangle,
\label{EQ89}
\end{align}
where $| 0_{nuR} \rangle$ is the non-uniform Rindler vacuum. We have found that the commutation relation for the non-uniform Rindler annihilation and creation operators is (\cite{MDAPHDTHESIS_2022})
\begin{align}
[\hat{c}_{\Omega}, \hat{c}_{\Omega}^\dagger]= 1,
\label{EQ90}
\end{align}
which, combined with Eq. \eqref{EQ88}, gives us, for $n_{k}$ even:
\begin{align}
\langle N_{nuR, n_{k}} | 0_{M} \rangle = \langle 0_{nuR} | 0_{M} \rangle \prod_{k} \big(G(\tau)\big)^{k} e^{\frac{-n_{k} \pi \Omega_{k}}{2a}} \frac{(n_{k} - 1)!!}{\big((n_{k})!\big)^{1/2}},
\label{EQ91}
\end{align}
and, for $n_{k}$ odd:
\begin{align}
\langle N_{nuR, n_{k}} | 0_{M} \rangle = 0.
\label{EQ92}
\end{align}
Hence, considering the expansion of the Minkowski vacuum in the Rindler-generalized eigenstates
\begin{align}
| 0_{M} \rangle = \sum_{n_{k}} | N_{nuR, n_{k}} \rangle \langle N_{nuR, n_{k}} | 0_{M} \rangle,
\label{EQ93}
\end{align}
we obtain the expression:
\begin{widetext}
\begin{align}
| 0_{M} \rangle = \langle 0_{nuR} | 0_{M} \rangle \sum_{m_{k}}\bigg( \prod_{k} \big(G(\tau)\big)^{k} e^{-m_{k} \pi \Omega_{k} / a} \frac{(2m_{k} - 1)!!}{\big((2m_{k})!\big)^{1/2}} \bigg) | N_{nuR, 2m_{k}} \rangle,
\label{EQ94}
\end{align}
\end{widetext}
where $n_{k}$ was replaced by $2 m_{k}$.
If we multiply this latter expression on the left with $| 0_{M} \rangle$ and use Eq. \eqref{EQ92}, then:
\begin{align}
& \langle 0_{M} | 0_{M} \rangle = \nonumber \\
& |\langle 0_{nuR} | 0_{M} \rangle|^2 \sum_{m_{k}}\bigg( \prod_{k} \big(G(\tau)\big)^{2k} e^{-2m_{k} \pi \Omega_{k} / a} \frac{\big((2m_{k} - 1)!!\big)^{2}}{(2m_{k})!} \bigg).
\label{EQ95}
\end{align}
It is straightforward to prove that
\begin{align}
\sum_{m_{k}}\bigg( e^{-2m_{k} \pi \Omega_{k} / a} \frac{\big((2m_{k} - 1)!!\big)^{2}}{(2m_{k})!} \bigg) = \big( 1 - e^{-2 \pi \Omega_{k} / a} \big)^{-1/2}.
\label{EQ96}
\end{align}
Hence, we define a new quantity $Q_{RG}$ as:
\begin{align}
Q_{nuR}^{-1} = |\langle 0_{nuR} | 0_{M} \rangle|^2 = \prod_{k} \big(G(\tau)\big)^{-2k} \big( 1 - e^{-2 \pi \Omega_{k} / a} \big)^{-1/2},
\label{EQ97}
\end{align}
which we substitute into Eq. \eqref{EQ96} and obtain:
\begin{align}
| 0_{M} \rangle = Q_{nuR}^{-\frac{1}{2}} \sum_{m_{k}}\bigg( \prod_{k} e^{-\frac{\pi m_{k} \Omega_{k}}{a}} \frac{(2m_{k} - 1)!!}{\big((2m_{k})!\big)^{1/2}} \bigg) | N_{nuR, 2m_{k}} \rangle.
\label{EQ98}
\end{align}
Again, it is straightforward to prove that
\begin{align}
&  \sum_{m_{k}}\bigg( e^{-m_{k} \pi \Omega_{k} / a} \frac{(2m_{k} - 1)!!}{\big((2m_{k})!\big)^{1/2}} \bigg) | N_{nuR, 2m_{k}} \rangle = \nonumber \\
& \exp{\Big(\frac{1}{2} e^{-\pi \Omega_{k} / a} (c_{\Omega, k}^\dagger)^{2} \Big)} | 0_{nuR} \rangle.
\label{EQ99}
\end{align}
Therefore, we finally obtain:
\begin{align}
| 0_{M} \rangle = Q_{nuR}^{-\frac{1}{2}} \bigg( \prod_{k} \exp{\Big(\frac{1}{2} e^{-\frac{\pi \Omega_{k}}{a}} (c_{\Omega, k}^\dagger)^{2} \Big)} \bigg) | 0_{nuR} \rangle,
\label{EQ100}
\end{align}
an expression for the Minkowski vacuum as a one-mode squeezed state.
%%%%%%%%%%%%%%%%%%%%%%%%%%%%%%%%%%%%%%%%%%%%%%%%%%%%%%%%%%%%%%%%%%%%%%%
%%%%%%%%%%%%%%%%%%%%%%%%%%%%%%%%%%%%%%%%%%%%%%%%%%%%%%%%%%%%%%%%%%%%%%%
%%%%%%%%%%%%%%%%%%%%%%%%%%%%%%%%%%%%%%%%%%%%%%%%%%%%%%%%%%%%%%%%%%%%%%%
\section{Conclusions}

In this work, we have successfully demonstrated that a non-inertial observer, equipped with our Unruh-DeWitt detector (\cite{AINGPBMDAXW_DADOUDWF_2020}), and undergoing non-uniform, time-dependent acceleration, perceives a particle distribution bearing a striking resemblance to the Planckian distribution characteristic of the standard Unruh effect in Rindler spacetime (\cite{UWG}). However, this distribution is uniquely characterized by a time-dependent factor, as evidenced by Eq. (\ref{EQ44}). This key finding not only aligns with the asymptotic conditions governing the problem but also offers a deeper understanding of the interplay between quantum field theory and non-inertial frames in more general settings.

Furthermore, our analysis reveals a profound implication for the nature of the quantum vacuum. We have rigorously established that the Minkowski vacuum, when viewed through the lens of an accelerating observer, undergoes a "deformation". This deformation can be mathematically represented in two equivalent ways: as an expansion of two-mode squeezed states in the Rindler basis, or as an expansion of one-mode squeezed states in the non-uniform Rindler basis. This crucial insight underscores the observer-dependent nature of particle content in quantum field theory and opens intriguing avenues for further exploration. The development of our modified Unruh-DeWitt detector (\cite{AINGPBMDAXW_DADOUDWF_2020, MDAPHDTHESIS_2022}) has paved the way for experimentally probing the Unruh effect under more realistic profiles, whether uniform or non-uniform accelerating ones. Our theoretical framework provides a crucial step towards bridging the gap between theoretical predictions and experimental verification of this fundamental quantum phenomenon.
%%%%%%%%%%%%%%%%%%%%%%%%%%%%%%%%%%%%%%%%%%%%%%%%%%%%%%%%%%%%%%%%%%%%%%%
\section*{Acknowledgments}
One of the co-authors, Prof. Alexander I. Nesterov, contributed substantially to the development and revision of this work, but passed away prior to its final completion. \\
Another co-author, Prof. Gennady P. Berman, passed away before the completion of this manuscript. His contributions to the ideas and conceptual development of this work are gratefully acknowledged.
%%%%%%%%%%%%%%%%%%%%%%%%%%%%%%%%%%%%%%%%%%%%%%%%%%%%%%%%%%%%%%%%%%%%%%%
\bibliographystyle{unsrt}
\bibliography{AM}

\end{document}